\documentclass[%
prd%
 ,twocolumn%
 ,secnumarabic%
,amssymb, amsmath,nobibnotes, aps, prl]{revtex4}
\usepackage{color}
\usepackage{graphicx}
\usepackage{multirow}

\begin{document}
\title{Suppression of parametric oscillatory instability in third generation gravitational wave detectors}
\author{S.E. Strigin}
\affiliation{M.V. Lomonosov Moscow State University, Faculty of Physics, Moscow 119991, Russia}
\email{strigin@phys.msu.ru}
%%% start of single column %%%
\begin{abstract}

We discuss one possible method for suppression  of the nonlinear effect of a parametric oscillatory instability in a Fabry-Perot cavity of the next generation of gravitational wave detectors  by determination of optimal radii of curvature of interferometers' mirrors. Varying the radii of curvature of the mirrors we  can change the frequencies of the optical modes of the Fabry-Perot resonators and distributions of optical fields in optical modes, which, in turn, can change the number of unstable modes, leading to parametric oscillatory instability. The determination of  the optimal values of the radii of curvature of mirrors to reduce the number of unstable modes by thermal  lensing is fulfilled. 

\end{abstract}
%%% end of single column %%%

\maketitle

\section{Introduction}

Currently, the European Gravitational Observatory is developing a project of a third generation gravitational wave interferometer that would allow solving more ambitious problems than the gravitational detectors such as LIGO, VIRGO, GEO and others. The future third generation instrumentation for searching for gravitational waves, which was named the Einstein Telescope(ET), will have a sensitivity several orders of magnitude higher than  that of second generation gravitational wave detectors. The details of this project are still in the development stage and the first data on the Fabry-Perot cavities that form the ET arms were outlined in \cite{et1,et2,et3}.     

Distance $L$ between the Fabry-Perot cavity mirrors will also be increased to 10km compared to 4km in LIGO, mirror radius will be increased to 31cm, which is about a factor of 4.25 larger than the laser-beam radius on the mirrors, thus making optical mode losses due to diffraction very small. To reduce the level of exposure to seismic influence, it is planned to place the third generation gravitational wave detector in a horizontal tunnel with two 10-km arms underground(see all other parameters of ET  in Table \ref{tab:props1}).

In full scale terrestrial interferometric gravitational wave antennae like ET the undesirable effect of parametric oscillatory instability (PI) in the Fabry-Perot (FP) cavities may take place\cite{bsv}. Due to the high optical power of up to $W\simeq 3$MW circulating in the arm cavities a transfer of optical energy driving mechanical oscillations of the mirrors is likely. This behaviour might disturb the locking performance of the interferometer and can cause a substantial decrease of the antennae sensitivity.

The effect of PI describes the excitation of the mirror's elastic modes fed by the optical energy in the resonator. Thus, the optical main mode (TEM$_{00}$) with frequency $\omega_0$ effectively drives the elastic motion of the mirror with frequency $\omega_m$ as well as a second optical mode with frequency $\omega_1$ which is called Stokes mode. 
Especially at the resonance $\omega_0-\omega_1=\omega_m$ the described energy transfer becomes very efficient. 
The coupling between these three modes arises both due to the ponderomotive pressure of light in the main and Stokes modes and the parametric action of mechanical oscillation on the optical modes. 
Above a critical value of optical power $W_c$ in the resonator the amplitude of the mirror's mechanical oscillation rises exponentially as well as the optical power in the Stokes mode.

In 2002 it was shown that the mirror's oscillation could be damped by the same mechanisms. 
So the mechanical oscillation of the mirror could drive an optical anti-Stokes mode with frequency $\omega_{a}=\omega_0+\omega_m$ resulting in the damping of the motion of the mirror\cite{ak}. 
This will of course lessen the negative consequences of PI for an interferometer.  
But in \cite{bsv2} it has been shown that the consideration of anti-Stokes modes could not completely suppress the effect of parametric oscillatory instability. 
This calculation was based on the geometry of the LIGO interferometer including the power recycling mirror. 
Also the detailed analysis of PIs for the LIGO interferometer including signal recycling has been provided in \cite{gras3,2007,gras1,gras2,gras4,gras5,gras6,2007_1}. 
 
The condition for the appearence of a parametric oscillatory instability for the FP cavity may
be written in a simple relation for the parametric gain ${\cal R}$ \cite{bsv}:
\begin{eqnarray} 
\label{equ:gain} {\cal R}=\frac{\Lambda_1 W\omega_1}{cLm\omega_m\gamma_m\gamma_1}\times
\frac{1}{1+\frac{\Delta^2}{\gamma_1^2}}>1 \ , \\
\label{equ:ovlapfactor} \Lambda_1=\frac{V(\int A_0(r_{\bot})A_1(r_{\bot})u_z dr_{\bot})^2}{\int|A_0|^2
dr_{\bot}\int|A_1|^2dr_{\bot}\int|\vec{u}|^2 dV} \ .
\end{eqnarray}
Here $c$ is the speed of light, $L$ is the length of the FP cavity, $m$ is the mirror's mass, 
$\gamma_m$ and $\gamma_1$ are the relaxation rates of the elastic and Stokes modes respectively, 
$W$ is the power circulating inside the cavity and $\Delta=\omega_0-\omega_1-\omega_m$ is the detuning value. 
$\Lambda_1$ characterises the overlapping factor between the elastic and the two optical modes (main mode and Stokes mode). 
$A_0$ and $A_1$ represent the optical field distributions at the reflecting surface of the mirror for the optical main mode and the Stokes mode, respectively. 
$\int d{r}_{\bot}$ corresponds to the integration over this reflecting surface and $\int dV$ -- over the mirror volume $V$.
The vector $\vec{u}$ describes the spatial displacement of the mirror's elastic mode and $u_z$ is the z-component of
$\vec{u}$ oriented along the cylindrical axis.

As we can see from Eq.~(\ref{equ:gain}) the parametric instability is a threshold effect and it takes place if the optical power $W$ in the main mode of the FP cavity is bigger than a threshold value $W_c$. 
The value $W_c$ depends on the detuning $\Delta$ between the optical and elastic modes and the overlapping factor $\Lambda_1$. 
In particular, $W_c$ takes its minimum value if the detuning is less than the relaxation rate of the Stokes mode $\Delta\ll \gamma_1$ and the overlapping factor is large enough.

Analyzing Eq.~(\ref{equ:gain}) the authors in \cite{ak} pointed out that the presence of the anti-Stokes mode may depress or even exclude PIs. 
%Indeed, including the effect of an anti-Stokes mode with frequency $\omega_{1a}$ the condition of PI has the following form \cite{bsv2}: 
Indeed, including the effect of an anti-Stokes mode with frequency $\omega_{1a}$ results in a negative gain
$${\cal R}_a= -\frac{W\omega_1}{cLm\omega_m\gamma_m\gamma_1} 
\times\frac{\omega_{1a}\gamma_1}{\omega_1\gamma_{1a}} \times
\frac{\Lambda_{1a}}{{1+\frac{\Delta_{a}^2}{\gamma_{1a}^2}}},$$
where $\gamma_{1a}$ is the relaxation rate of the anti-Stokes mode and $\Delta_{a}=\omega_{1a}-\omega_0-\omega_m$ is the detuning value for the anti-Stokes mode.
This damping effect has to be considered in a PI search. The effective parametric gain for each elastic mode has the following form ${\cal R}_{tot}={\cal R}+{\cal R}_a$.

In particular, we need detailed information about pairs of Stokes modes and elastic modes which may be
possible candidates for parametric instability. 
It is known that Stokes and anti-Stokes modes may be analytically calculated for Gaussian beams.  
In contrast, elastic modes can only be calculated numerically. 

Imperfections of the mirrors from the cylindrical shape (such as flats and suspension
'ears') cause both splitting of originally mechanical degenerate modes into doublets and  an additional frequency shifts of the elastic modes.  If, in turn, the difference between doublet frequencies is large enough, i.e. greater than the relaxation rate of the Stokes mode new PIs could emerge. Thus, the appearance of doublets may increase the possibility of PI \cite{2008,2008_1}.

In order to predict parametric unstable combinations of Stokes and elastic modes we use an additional azimuthal condition. 
Let us assume TEM$_{00}$ as the optical main mode. 
If the elastic mode and the Stokes mode show an azimuthal dependence of $\sim e^{i l\phi}$ and $\sim e^{i m\phi}$, respectively, the overlapping factor will vanish unless $m=l$. 
This behaviour results from the azimuthal integration in Eq.~(\ref{equ:gain}) for $\Lambda_1$.
It is worth noting that this is correct only if the cylinder center and the laser spot center coincide. For Laguerre-Gaussian optical mode LG$_{33}$ which uses as a pump in ET the azimuthal condition is $|m \pm l|=3.$
 
In this Letter we analyse PI in ET at different values of radii of curvature and find  optimal ones  for effective suppression of PI.
In section I we introduce all details of our PI calculations. In section II we deduce the dependence of the number of unstable modes on radius of curvature(ROC) for the end test mass(ETM) and find optimal range of its ROCs. The ROC of input test mass(ITM) is fixed as in Table \ref{tab:props1}.

\section{I. PI in ET}

%\subsection{Elastic modes}

The most important requirement for a PI search consists in the knowledge of the mechanical resonant frequencies and their elastic mode shapes. 
For elastically isotropic and cylindrical substrates there exist analytical solutions specialised to some restrictive conditions.
One example is the well-known Chree-Lamb modes \cite{chree,lamb} that are only applicable for certain ratios of the cylinder's radius  and height   and show no azimuthal deformation.
Recent investigations using the superposition method allow to calculate the resonant frequencies and mode shapes of all axially symmetric modes of a cylinder with high accuracy. 
In their work Meleshko et al. \cite{meleshko} derived partial solutions of the equation of motion that satisfy the boundary condition of the free sample.  But the only method to obtain the complete set of elastic mode shapes and frequencies for our PI calculation consists in a finite element modelling. 
For this purpose the FE package COMSOL$^\circledR$ was used. In this Letter we propose accuracy estimates to be about 1\% due to analytical solutions of Chree-Lamb-modes.

%\begin{table}
%\begin{tabular} {ccccc}
%set $\#$ & R [m] &H in [m] & $f_{ana}$ [Hz] & comments \\
%1 & 0.193& 0.155 & 51433 & $a_4$ and $b_2$ \\
%2	& 0.214&0.126 &21110& $a_2$ and $b_1$ \\
%3	& 0.17& 0.2 & 18105 & $c_1$ and $d_0$ \\
%\end{tabular}
%\caption{Test cases for estimate of numerical errors in the FE codes. The comment column shows the elastic mode index fully identifying %the mode under consideration. The sample dimension of the Chree-Lamb modes were chosen in a way to keep the sample volume constant %\textcolor{red}{at a value equivalent to the proposed 40kg aLIGO mirror.}}
%\label{tab:testmodes}
%\end{table}

For the mechanical relaxation rate $\gamma_m$ we use the model of structural loss and express the relaxation rate with respect to mechanical loss $\phi$. 
Within this approximation we find a frequency dependent mechanical relaxation rate
\begin{align}
\gamma_m\approx\frac{1}{2}\omega_m\phi \ .
\label{equ:gammamech}
\end{align}
Note that this approximation is only valid for small mechanical loss values.
In the further calculation we use a constant value of $\phi=1\times10^{-8}$ for fused silica at room temperature.

%\subsection{Optical modes}
%\label{sec:optmodes}

Optical resonant frequencies and the spatial distribution of their electric field on the surface of the mirror define the possibility of PI. 
The spectrum of optical mode frequencies depends on the geometry of the FP cavity, which is realised in the interferometer arms of ET.

The optical mode frequencies in a Fabry-Perot cavity can be calculated due to Kogelnik \cite{kogelnik} as
\begin{align}
\omega_{\rm Nnm}=\frac{c}{L}\left(N_0 \pi+(2n+m+1)\times \phi_{0}\right)\ ,\nonumber\\ 
\phi_0=\arccos{\left[- \sqrt{\left(1-\frac{L}{R_1}\right)\times\left(1-\frac{L}{R_2}\right)}\right]}\ .
\label{eq1}
\end{align} 
There $N_0 = 0,\pm 1,\pm 2,\ldots$ is the longitudinal index, $n = 0, 1, 2,\ldots$ is the radial index and $m = 0, 1, 2,\ldots$ is the azimuthal index.
$R_1$ and $R_2$ are the radii of curvature for the ITM and ETM of the Fabry-Perot cavity(see details in Table \ref{tab:props1}).
The optical main mode circulating inside the cavity in ET is identified by $n=3$ and $m=3$($LG_{33}$ mode).

Consequently, the transition from the optical main mode $\omega_0$ to another optical mode $\omega_1$ is accompanied by a frequency change of
\begin{align}
\Delta\omega=\omega_1-\omega_0=\frac{c}{L} \left[(2n+m)\times \phi_0-\pi N \right] \ .
\label{equ:deltaf}
\end{align}
Note that only negative frequency changes $\Delta\omega<0$ can cause PIs as they belong to an optical Stokes mode. 
In contrast positive frequency changes belong to anti-Stokes modes, which can effectively suppress PIs and will be considered in a separate calculation.

It is worth noting that not only the frequency detuning $\Delta$ but also the loss of optical energy influences the parametric gain. The latter is summarised in the optical relaxation rate $\gamma_1$.
On the one hand, energy loss occurs due to transmission of the mirrors  and, on the other hand,  due to clipping losses as a result of diffraction . 
For our calculation we only consider the coefficient of energy transmission for the input test mass $T_{\rm ITM}$ as well as clipping energy losses on both mirrors $T_{clip}$. Both these coefficients create effective transmission coefficient $T_{tot}$\cite{hei}.

This calculation shows an increase of optical loss for increasing mode orders due to the clipping loss fraction. 
Thus, the effect of parametric instability of high order optical modes becomes more and more unlikely (see $\gamma_1$ in Eq.~(\ref{equ:gain})). 
In this work we only considered optical modes whose transversal orders fulfill the condition $2n+m\leq 10$.

\subsection{ II. RESULTS}

The calculation of parametric gain ${\cal R}$ follows from Eqs.~(\ref{equ:gain}), (\ref{equ:gammamech}) and (\ref{eq1}). 
As the elastic frequencies are orders of magnitude smaller than the optical ones we can use the relation $\omega_1=\omega_0-\omega_m\simeq \omega_0$. 
Together with the dispersion relation of light and the wavelength of the optical main mode $\lambda$
\begin{align}
 \omega_1/c \simeq \omega_0/c = 2\pi/\lambda 
\end{align}
we can simplify the expression for the parametric gain to
\begin{align}
{\cal R} \simeq \frac{16 \pi}{c} \frac{W}{\lambda m \phi} \frac{1}{\omega_m^2 T_{tot}}
\times \frac{\Lambda_1}{1+\frac{\Delta^2}{\gamma_1^2}} \ .
\end{align}

Using a MATLAB$^\circledR$ code we evaluated this expression for mechanical modes up to a resonant elastic frequency of 30\,kHz and for  optical Stokes modes confined to mode orders of $2n+m\leq 10$.
With these numbers we obtain a total of nine possible PI candidates in ET for the mentioned range of optical and mechanical modes. 

By the same time the mirror is excited by a PI including a Stokes mode it could as well be damped by the effect of PI due to feeding of an anti-Stokes mode. Subsequently the gain for the mirror oscillation could fall below the threshold of ${\cal R}=1$  and prevent parametric instability to occur for the given oscillation.

In a detailed analysis we took the elastic mode of each candidate and calculated the parametric gain ${\cal R}_a$ of the anti-Stokes effect for optical modes satisfying $2n+m\leq 10$. Finally we added all anti-Stokes gains ${\cal R}_a$ and subtracted this value from the original parametric gain value ${\cal R}$. This procedure lead to a total of nine  parametric unstable combinations. 

\begin{figure}[ht]
	\begin{center}
		\includegraphics[width=8
cm,totalheight=7cm]{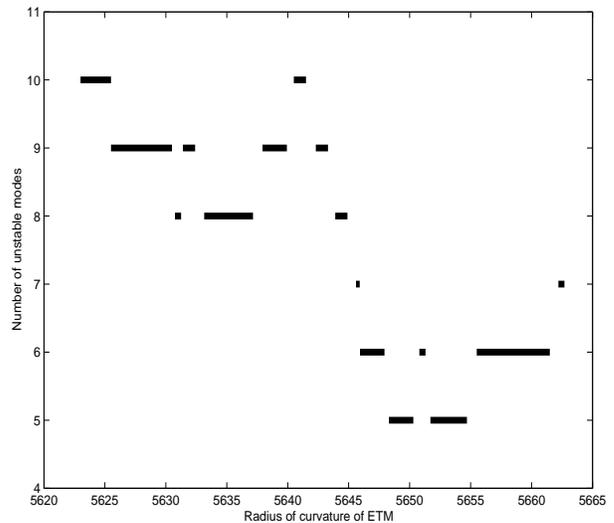}
		\caption{The number of unstable elastic modes  with ${\cal R}_{tot}>1$ while tuning the radius of curvature of ETM}	
		\label{fig:diagr}
	\end{center}
\end{figure}

One possible method to minimize the number of unstable modes is the changing of mirrors' radii of curvature by local heating using additional laser(thermal lensing). This idea was firstly pointed out in \cite{gras3,gras1,gras7} for LIGO interferometer. The thermal tuning of optical Fabry-Perot cavities has been proposed to reduce the parametric gain of parametric instabilities\cite{gras7}. The authors investigated the performance achievable for such tuning obtained by thermal actuation of the mirrors of the arm cavities.  In our Letter we present the similar method for reduction of PIs and propose to vary the radius of curvature of ETM in one Fabry-Perot cavity. The range of these variations is from 5623 to 5663 meters. We took this interval into account  because of making small changes in ET design.

In Fig.\ref{fig:diagr} the dependence of number of unstable modes on the radius of curvature of ETM is plotted. In our calculations we assume that each non-axisymmetric elastic mode has sine and cosine component in azimuthal angle with practically the same frequencies. Therefore, each component gives an equal contribution to the parametric gain ${\cal R}_{tot}$. In Fig.\ref{fig:diagr} we consider each unstable elastic mode as one unstable mode having these two components.  It is worth noting that the planed value of radius of curvature for ETM($R_2=5643$m) is not optimal one because of large number of unstable modes(nine unstable modes only in frequency range of elastic modes up to 30kHz). Varying the radius of curvature of ETM  $R_2$, for example, by thermal lensing(the $R_1$ is a constant one) we can suppress the PI approximately by a factor of 2(five unstable modes) if the radius of curvature of ETM mirror becomes slightly bigger. The prefered values of possible radii of curvatures lie in two ranges from 5648.3 to 5650.8 meters and from 5651.7 to 5655.5 meters. Because of some inaccuracy in elastic parameters and  numeric calculations the better interval for suppression of PI is a second long interval.  In Table \ref{tab:addlabel} all unstable combinations of elastic and optical modes with appropriate parametric gains are shown(see Appendix A). If we do not take into account the influence of modes splitting some elastic modes will be stable. For example, for the ROC of ETM $R_2=5623$m elastic modes with the frequencies $17091$Hz and $25339$Hz become stable(for $R_2= 5633$m --  $20285$Hz; for $R_2=5643$m -- $13572$Hz, $13701$Hz, $25339$Hz; for $5653$m -- $23387$Hz; for $5663$m -- $28398$Hz correspondingly).  

\section{Conclusion}
In this Letter we have analyzed the possibility of the reduction of the unstable modes in ET by varying, for example, the radius of curvature of ETM mirror. In this case the frequencies of the optical modes of the Fabry-Perot resonators and distribution of their optical fields on the mirrors' surfaces change, which, in turn, can change the number of unstable modes leading to parametric oscillatory instability.

Another important problem related to PI is the increase in the elastic modes' density for high frequencies which is caused by deviations of the shape of real test masses of the ET from an ideally cylindrical shape(they have ''ears'' for suspension and symmetrically arranged flat cuts on the sides of the cylinders\cite{2008,2008_1}). Such changes of the mirror shape result in a shift of the frequency spectrum of elastic modes and splitting of non-axisymmetric modes into doublets which increase the density of elastic modes by nearly twofold. These factors, in turn, stimulate high probability of the onset of the PI in ET and change the conditions during which the PI is observed.

It is worth noting that the mean interval between neighboring elastic modes near the elastic frequency $\omega_m = 2\pi \times 30000$Hz can be estimated as  $\Delta \omega_m \simeq 100 \mbox{s}^{-1}$\cite{bsv,bsv2}. But in real detectors the mirrors have flats and suspension ''ears'' which increase the number of mechanical modes and, in turn, the probability of PI. A value of splitting gap between sine and cosine components for non-axisymmetric elastic modes can archive $\delta \omega_m \simeq 40\mbox{s}^{-1} $. Neglecting the clipping losses(for optical modes with $2n+m\leq 10$ these losses are negligible) we find relaxation rate of these modes to be $\gamma_1 \simeq 55\mbox{s}^{-1}$. Therefore, for elastic modes near the elastic frequency $\omega_m = 2\pi \times 30000$Hz(taking into account the splitting) the mean interval between neighboring elastic modes is approximately equal to the relaxation rates of the optical modes  $\Delta \omega_m/2 \simeq \gamma_1$. It means that elastic spectrum becomes continious and practically for each elastic mode with frequency higher than 30kHz there exists the Stokes mode with small detuning $\Delta < \gamma_1$. For this elastic modes range the number of unstable modes will be practically the same for different radii of curvature. But only accurate calculation of overlapping factors is needed for searching  of unstable elastic modes.

Our results have  stochastical character because of variations in elastic parameters(Youngs modulus, density and others). Therefore, the values of radii of curvature can change and influence the number of possible PIs.  On the other hand,  the presence of power
recycling cavity causes parametric gains ${\cal R}_{tot}$ to be hugely amplified\cite{bsv2}.  In this case the experimental results are needed.

\section*{Acknowledgement}
The author kindly acknowledges the support of the Russian Foundation for Basic Research(grants nos. 11-02-00383-a and 14-02-00399-a). 

\begin{appendix}

\section{A. Parameters and detailed results of the calculations}
\label{app:params}
In this Appendix we present the numerical values of the parameters of ET that influence our calculations. 
As the substrates are made from fused silica we used material parameters according to this material at room temperature. 
The numerical values are presented in Table~\ref{tab:props1}. The parametric gains for some values of radius of curvature for ETM  are shown in Table~\ref{tab:addlabel}.  

\begin{table}
\begin{tabular}{|lc|}
\multicolumn{2}{l}{elastic modes}\\
\hline
ETM radius 								& 31\,cm \\
ETM height									&	30\,cm \\
Young's modulus $Y$					&73.1\,GPa \\
Poisson's ratio $\sigma$		&0.17 \\
density $\rho$							&2203\,kg/m$^3$ \\
\hline
\multicolumn{2}{l}{optical modes} \\
\hline
\multicolumn{2}{|l|}{radius of curvature}							\\
ITM	$R_1$												&5643\,m \\
ETM	$R_2$												&5643\,m \\
\hline
\multicolumn{2}{|l|}{substrate diameter}	\\	
ITM													&62\,cm \\
ETM													&62\,cm		\\
\hline
\multicolumn{2}{|l|}{laser beam radius $w$} \\
ITM													&7.3\,cm \\
ETM													&7.3\,cm 	\\
\hline
resonator length $L$						&	10000\,m \\
\hline
\multicolumn{2}{l}{parametric gain}\\
\hline
\multicolumn{2}{|l|}{energy transmission}		\\
ITM	$T_\mathrm{ITM}$				&	$7.36\times10^{-3}$			\\
ETM													& 0	\\
\hline
mechanical loss $\phi$				&$1\times10^{-8}$ \\
laser wavelength $\lambda$	& 1064\,nm \\
mirror mass (ETM) $m$						&200 \,kg \\
optical power inside cavity $W$	&3\,MW \\
speed of light $c$					&$3\times 10^{8}$\,m/s \\
\hline
\end{tabular}
\caption{Numerical values of the parameters for the calculation of parametric instabilities.}
\label{tab:props1}
\end{table}

%\section{Results}
%\subsection{List of PIs}
%\label{app:resultsPI}

\begin{table}[htbp]
  \centering
    \begin{tabular}{cccccccc}
    \ $R_2 [m]$  & $R_1$ [m] & $f_m$ [Hz]	& $l$  & $n$     & $m$     & ${\cal R}_{tot}$   \\
    \hline
    5623   & 5643 & 8486 & 0     & 2     & 3  & 370.17   \\
    5623   & 5643 & 8764 & 0     & 2     & 3  & 4.6   \\
     5623   & 5643 & 13701 & 1     & 1     & 2  & 50   \\
      5623   & 5643 & 13701 & 1     & 0     & 4  & 6.66   \\
       5623   & 5643 & 16913 & 2     & 2     & 1  & 123.21   \\
        5623   & 5643 & 17091 & 2     & 2     & 1  & 1.17   \\
        5623   & 5643 & 20194 & 1     & 1     & 4  & 32.6   \\
    5623   & 5643 & 20194 & 1     & 2     & 2  & 7.52   \\
     5623   & 5643 & 20285 & 1     & 2     & 2  & 2.12   \\
      5623   & 5643 & 25339 & 2     & 1     & 1  & 1.3   \\
       5623   & 5643 & 25394 & 2     & 1     & 1  & 3   \\
        5623   & 5643 & 28682 & 1     & 0     & 4  & 154.26   \\
          5623   & 5643 & 28682 & 1     & 1     & 2  & 1.15   \\
    \hline
   5633   & 5643 & 8486 & 0     & 2     & 3  & 70.61   \\
    5633   & 5643 & 8764 & 0     & 2     & 3  & 3.93   \\
     5633   & 5643 & 13701 & 1     & 1     & 2  & 4.39   \\
       5633   & 5643 & 16913 & 2     & 2     & 1  & 292.1   \\
        5633   & 5643 & 20194 & 1     & 1     & 4  & 13.55   \\
    5633   & 5643 & 20194 & 1     & 2     & 2  & 3.13   \\
      5623   & 5643 & 20285 & 1     & 2     & 2  & 1.01   \\
     5633   & 5643 & 23368 & 0     & 2     & 3  & 1.1   \\
      5633   & 5643 & 25339 & 2     & 1     & 1  & 81.81   \\
         5623   & 5643 & 28682 & 1     & 0     & 4  & 3.46  \\
         \hline
   5643   & 5643 & 8486 & 0     & 2     & 3  & 28.05  \\
    5643   & 5643 & 8764 & 0     & 2     & 3  & 3.4   \\
    5643   & 5643 & 12096 & 1     & 3     & 2  & 1.7   \\
        5643   & 5643 & 13572 & 1     & 1     & 2  & 1.5   \\
       5643   & 5643 & 16913 & 2     & 2     & 1  & 26.97   \\
        5643   & 5643 & 20194 & 1     & 1     & 4  & 2.31   \\
   5643   & 5643 & 21967 & 3     & 1     & 0  & 4.25   \\
     5643   & 5643 & 23368 & 0     & 2     & 3  & 2.22   \\
      5643   & 5643 & 25339 & 2     & 1     & 1  & 1.62   \\
          \hline
   5653   & 5643 & 8486 & 0     & 2     & 3  & 14.92  \\
    5653   & 5643 & 8764 & 0     & 2     & 3  & 2.98  \\
       5653   & 5643 & 16913 & 2     & 2     & 1  & 9.21   \\ 
       5653   & 5643 & 23368 & 0     & 2     & 3  & 6.46  \\
       5653   & 5643 & 23387 & 10     & 0     & 7  & 1.67   \\ 
           \hline
   5663   & 5643 & 8486 & 0     & 2     & 3  & 9.24  \\
    5663   & 5643 & 8764 & 0     & 2     & 3  & 2.63   \\
       5663   & 5643 & 16913 & 2     & 2     & 1  & 4.6   \\
        5663   & 5643 & 23368 & 0     & 2     & 3  & 48.63   \\
   5663   & 5643 & 23387 & 10     & 0     & 7  & 2.24   \\
      5663   & 5643 & 28411 & 3     & 2     & 0  & 5.43   \\
   5663   & 5643 & 28398 & 1     & 1     & 2  & 1.22   \\
    \hline
    \end{tabular}%
    \caption{Candidates for parametric instabilities. Given are the  mirrors' radii of curvature of Fabry-Perot cavity, the elastic resonant frequency $f_m=\omega_m/(2\pi)$, azimuthal number of elastic mode ($l$), the radial ($n$) and azimuthal ($m$) indeces of the optical mode and the effective  parametric gain ${\cal R}_{tot}={\cal R}+{\cal R}_{a}$. The contribution of possible anti Stokes modes is considered in the calculations.}
  \label{tab:addlabel}%
\end{table}%

\end{appendix}

\bibliographystyle{iopart-num}
\bibliography{refs}

\end{document}